\documentstyle[sprocl,epsfig]{article}

\def\Journal#1#2#3#4{{#1} {\bf #2}, #3 (#4)}

\def\NPB{{\em Nucl. Phys.} B}
\def\PLB{{\em Phys. Lett.}  B}

\def\PRD{{\em Phys. Rev.} D}
\def\ZPC{{\em Z. Phys.} C}
\def\PREP{{\em Phys. Rep.}}
\def\frac#1#2{ {{#1} \over {#2} }}

\def\half{\mbox{\small $\frac{1}{2}$}}

\def\abs#1{\left| \: #1 \: \right|}%
\def\bom#1{\mbox{\boldmath$#1$}}
\def\beq{\begin{equation}}
\def\eeq{\end{equation}}
\def\re#1{(\ref{#1})}
\def\ee{$e^+e^-\;$}
\def\as{\alpha_S}
\def\asb{\bar \alpha_S}

\def\bk{\bom {k} }
\def\bq{\bom {q} }
\def\bkq{\abs{\bom{k}+\bom{q}}}
\def\om{\omega}
\def\ga{\gamma}
\def\tga{\tilde \gamma}
\def\tchi{\tilde \chi}

\def\de{\delta}
\def\De{\Delta}
\def\cF{{\cal F}}
\def\cA{{\cal A}}
\def\Q_s{\mu}

\def\thefootnote{\dagger}

\begin{document}

\title{EVOLUTION EQUATIONS AND ANGULAR ORDERING AT SMALL 
$\bom{x}$ \footnote{Talk presented at the ``Conference on
Perspectives in Hadronic Physics'', Trieste, 12-16 May 1997}}

\author{M. SCORLETTI}

\address{Dipartimento di Fisica, Universit\`a di Milano
and INFN, sezione di Milano\\
via Celoria 16, 20133 Milano, ITALY}

\begin{flushright}
IFUM 574-FT \\
hep-ph/9707237\\[12pt]
\end{flushright}

\maketitle\abstracts{
This talks examines the effect of angular ordering on the small-$x$
evolution of the unintegrated gluon distribution, and discusses the
characteristic function for the CCFM equation.}
  
\def\thefootnote{\alph{footnote}}%
\setcounter{footnote}{0}%

\section{Introduction}

Angular ordering is an important feature of perturbative QCD \cite{IR}
with a deep theoretical origin and many phenomenological 
consequences.\cite{AngOrd}  
It is the result of destructive interference: outside
angular ordered regions amplitudes involving soft gluons cancel.
This property is quite general, and it is present in both time-like
processes, such as \ee annihilation, and in space-like processes, such
as deep inelastic scattering (DIS).  
Moreover it is valid in the
regions both of large and small $x$, in which $x$ is the registered
energy fraction in the \ee fragmentation function or the Bjorken
variable in the DIS structure function.  
Due to the universality of
angular ordering one has a unified leading order description of all
hard processes involving coherent soft gluon emission.

The detailed analysis of angular ordering in
multi-parton emission at small $x$ and in the related virtual corrections
\cite{CCFM,March} 
shows that to leading order the initial-state gluon emission can be
formulated as a branching process in which angular ordering 
is taken into account both in real emissions   
and virtual corrections. 

In DIS, angular ordering is essential for
describing the structure of the final state, but not for the gluon
density at small $x$.  
This is because in the resummation of singular
terms of the gluon density, there is a cancellation between the real
and virtual contributions. 
As a result, to leading order the small-$x$ gluon density is obtained 
by resumming $\ln x$ powers coming only from IR singularities, and angular
ordering contributes only to subleading corrections.

The calculation of the gluon density by resummation of $\ln x$ powers
without angular ordering was done $20$ years ago \cite{BFKL} and
leads to the BFKL equation, which is an evolution equation for 
$\cF(x,k)$, the unintegrated gluon density at fixed transverse momentum $k$:
\begin{equation}
  \label{bfkl}
  x {\partial \cF(x,k) \over \partial x} = 
  \asb \int \frac{d^2 \bq}{\pi q^2} 
  [ \cF(x,\bkq) - \theta(k-q) \cF(x,k) ]
\end{equation}
where $\asb = {C_A \over \pi} \as$. 
$\cF(x,k)$ is related to the small-$x$ part of the
gluon structure function $F(x,Q)$ by
\begin{equation}\label{cF}
F(x,Q)=\int d^2\bk \;\cF(x,k)\theta(Q-k)\,.
\end{equation}

In this talk,
as a first step of a systematic study of multi-parton emission in DIS, 
the effect of angular ordering on the small-$x$ evolution of the gluon 
structure function is studied,\cite{bmss} with both analytical and numerical
techniques. 

\section{Evolution equation for gluon density}

In this section we recall the basic ingredients used to build the
coherent branching equation for the gluon density at small $x$.  

\begin{figure}
\begin{center}
\begin{picture}(0,0)%
\epsfig{file=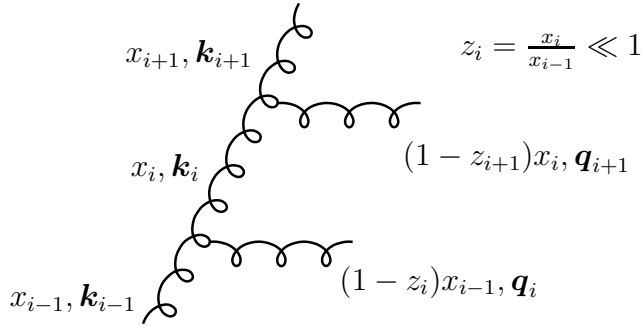}%
\end{picture}%
\setlength{\unitlength}{0.00087500in}%
\begingroup\makeatletter\ifx\SetFigFont\undefined
\def\x#1#2#3#4#5#6#7\relax{\def\x{#1#2#3#4#5#6}}%
\expandafter\x\fmtname xxxxxx\relax \def\y{splain}%
\ifx\x\y   
\gdef\SetFigFont#1#2#3{%
  \ifnum #1<17\tiny\else \ifnum #1<20\small\else
  \ifnum #1<24\normalsize\else \ifnum #1<29\large\else
  \ifnum #1<34\Large\else \ifnum #1<41\LARGE\else
     \huge\fi\fi\fi\fi\fi\fi
  \csname #3\endcsname}%
\else
\gdef\SetFigFont#1#2#3{\begingroup
  \count@#1\relax \ifnum 25<\count@\count@25\fi
  \def\x{\endgroup\@setsize\SetFigFont{#2pt}}%
  \expandafter\x
    \csname \romannumeral\the\count@ pt\expandafter\endcsname
    \csname @\romannumeral\the\count@ pt\endcsname
  \csname #3\endcsname}%
\fi
\fi\endgroup
\begin{picture}(3443,1948)(720,-1523)
\put(2213,-1396){\makebox(0,0)[rb]{\smash{\SetFigFont{12}{14.4}{rm}$x_{i-1},\bk_{i-1}$}}}
\put(2595,-616){\makebox(0,0)[rb]{\smash{\SetFigFont{12}{14.4}{rm}$x_{i},\bk_{i}$}}}
\put(2911, 74){\makebox(0,0)[rb]{\smash{\SetFigFont{12}{14.4}{rm}$x_{i+1},\bk_{i+1}$}}}
\put(3818,-548){\makebox(0,0)[lb]{\smash{\SetFigFont{12}{14.4}{rm}$(1-z_{i+1})x_{i},\bq_{i+1}$}}}
\put(3436,-1298){\makebox(0,0)[lb]{\smash{\SetFigFont{12}{14.4}{rm}$(1-z_i)x_{i-1},\bq_{i}$}}}
\put(4163,112){\makebox(0,0)[lb]{\smash{\SetFigFont{12}{14.4}{rm}$z_i={x_i\over
x_{i-1}}\ll1$}}}
\end{picture}
\end{center}
\caption[]{Labelling of momenta and approximate kinematic in 
the diagram for a DIS
process at parton level: $x_i$ and $\bk_i$ denote the energy
fraction and the transverse momentum of the $i$-th transmitted
gluon, while $\bq_i$ is the transverse momentum of the $i$-th
emitted gluon.}
\label{kine}
\end{figure}

The evolution of the gluon density can be described (Fig.~\ref{kine})
as a multi-branching process
involving only gluons, since gluons dominate the small-$x$ region.
The emission process takes place in the angular ordered region given
by $\theta_{i}>\theta_{i-1}$ with $\theta_{i}$ the angle of the
emitted gluon $q_i$ with respect to the incoming gluon $k_{0}$.
In terms of the emitted transverse momenta $q_i$ this region
is given by
\begin{equation}\label{ao}
\theta_{i}>\theta_{i-1}\,,
\;\;\;\;\;\;\Rightarrow\;\;\;\;\;\;
q_{i} > z_{i-1} q_{i-1}
\,.
\end{equation}
The branching distribution for the emission of gluon $i$ reads
\begin{equation}\label{dP}
d{\cal P}_i
=\frac{d^2\bq_i}{\pi q_i^2} \; dz_i\frac{\asb}{z_i}
\;\De(z_i,q_i,k_i)\;\theta(q_i-z_{i-1}q_{i-1})
\,,
\end{equation}
where 
\begin{equation}\label{De}
\ln \De(z_i,q_i,k_i)=
-\int_{z_i}^1 dz' \;\frac{\asb}{z'}
\int\frac{dq'^2}{q'^2}\;\theta(k_i-q')\;\theta(q'-z'q_i)
\,
\end{equation}
is the form factor which resums
important virtual corrections for 
small $z_i$.\footnote{The usual Sudakov form factor is not included in the
single-branching kernel, since it is cancelled by soft emissions.}
The branching \re{dP} --- which includes angular ordering \re{ao} both
in the real and the virtual emissions --- 
is accurate to leading IR order.\cite{CCFM,CFMO}

The ``non-Sudakov'' form factor \re{De} has a simple probabilistic 
interpretation. 
It corresponds to the probability for having no radiation of gluons with
energy fraction $x'=z'x_{i-1}$ in 
the region $x_i<x'<x_{i-1}$, and with a  transverse
momentum $q'$ smaller than the total emitted transverse momentum $k_i$
and with an angle $\theta'>\theta_i$. 
The two boundaries in $q'$ are due to coherence in the exchanged gluon
($k>q'$) and in the emitted one ($\theta'>\theta_i \ \Rightarrow\ q'>z'q_i$). 

Angular ordering provides a lower bound on transverse momenta, so that no
collinear cutoff is needed other than a small virtuality fot the first
incoming gluon. 
On the other hand, in order to deduce a recurrence
relation for the inclusive distribution
in the last gluon with fixed $x=x_n$ and $k= k_n$ one has to
introduce an additional dependence on a momentum variable $p$. 
That variable corresponds to the transverse momentum associated with
the maximum
available angle $\bar\theta$ for the last emission, which in DIS is
settled by the angle of the quarks produced in the boson-gluon fusion.
The dependence on $p$ is through 
\begin{equation}\label{max}
\theta_n <\bar \theta
\;\;\;\;\;\;\Rightarrow\;\;\;\;\;
z_nq_n < p 
\,,
\end{equation}
where $p \simeq xE\bar\theta$ and $xE$ is the energy of the $n$-th gluon, 
which undergoes the hard collision at the scale $Q$. 

The distribution for emitting $n$ initial state gluons is defined as 
\begin{equation}\label{An}
\cA^{(n)} (x,k,p) \;=\; \int \prod_{i=1}^{n} \; d {\cal P}_i \;
\theta(p-z_nq_n) \; \de(k^2-k_n^2) \; \de(x-x_n)
\,,
\end{equation}
so that the fully inclusive gluon density
\begin{equation}\label{A}
\cA(x,k,p) \;=\; \sum_{n=0}^{\infty} \; \cA^{(n)} (x,k,p)
\,,
\end{equation}
satisfies the equation (CCFM equation \cite{CCFM})
\begin{equation}\label{A1}
\cA(x,k,p) = \cA^{(0)}(x,k,p) +
\int\frac{d^2\bq}{\pi q^2} \frac{dz}{z}
\;\frac{\asb}{z}\De(z,q,k)
\theta(p-zq)\;\cA\left({x\over z},\bkq,q\right)
\,,
\end{equation}
where the inhomogeneous term $ \cA^{(0)}(x,k,p) $ is the
distribution for no gluon emission.

It can be proved \cite{bmss} that the gluon density $\cA(x,k,p)$ 
becomes independent of $p$ for $p\to\infty$.
Indeed, neglecting the $p$-dependence in $\cA(x,k,p)$ corresponds
to neglecting angular ordering.
In this case the transverse momenta have no lower bound, and we need
to introduce a collinear cutoff $\mu$ to avoid singularities.
We then modify the branching distribution in \re{dP}
and the virtual corrections \re{De} by the substitution
$\theta(q_i-z_{i-1}q_{i-1}) \to \theta(q_i-\mu)$
and $\theta(q'-z'q) \to \theta(q'-\mu)$ respectively.
The modified branching distribution reads
\begin{equation}\label{dP0}
d{\cal P}_i^{(0)}
=\frac{d^2\bq_i}{\pi q_i^2} \; dz_i\frac{\asb}{z_i}
\;\De^{(0)} (z_i,k_i)\;\theta(q_i-\mu)
\,,
\end{equation}
with the form factor
\begin{equation}\label{dD0}
\ln \De^{(0)}(z,k)=
-\int_z^1 dz' \;\frac{\asb}{z'}
\int\frac{dq'^2}{q'^2}\;\theta(k-q')\;\theta(q'-\mu)
\,.
\end{equation}

The gluon density $\cF(x,k)$
(in this case there is no dependence on the ``maximum angle'' $p$)
satisfies the following recurrence relation:
\begin{equation}\label{bfkl1}
\cF(x,k) \;=\; \cF^{(0)}(x,k) \;+\;
\int\frac{d^2\bq}{\pi q^2}\; \frac{dz}{z}
\;\frac{\asb}{z}\De^{(0)}(z,k)\;\theta(q-\mu)\;\cF\left({x\over z},\bkq\right)
\,.
\end{equation}

Although every branching factor \re{dP0} is divergent in the limit
of vanishing $\mu$, collinear singularities cancel in the inclusive
sum which defines the structure function. 
As a consequence, the limit $\mu\to 0$ can be safely performed 
in \re{bfkl1} which --- in this limit --- prove to be equivalent to the
BFKL equation \re{bfkl}.
In spite of the very different behaviour 
in the collinear region of the branching distributions \re{dP} and \re{dP0},
the neglecting of angular ordering has no effect on the 
structure functions at leading order.

This is no longer true for exclusive quantities.
Here collinear singularities do not cancel any more, and angular ordering 
becomes essential to control the structure of singularities.
A fixed cutoff $\mu$ regulates the collinear divergence which is present, 
but gives the wrong final state properties. 
The elimination of a large fraction of the small-transverse-momentum
emissions indeed means that angular ordering has a big effect on the final
state. 

\subsection{Properties of gluon distributions}

The major task of this talk is to examine the 
corrections to structure function evolution
that arise from angular ordering, which is  expected to be part
of the full NLO contribution.\cite{NLO}

As is well known, the BFKL equation has eigensolutions 
(strictly speaking eigensolutions of the equation without an 
inhomogeneous term and with no upper limit in the $z$ integral) 
of the form:
\begin{equation}
  \label{bfkl-eigen}
  x\cF(x,k) = 
  x^{-\om} \frac{1}{k^2}\left(\frac{k^2}{k_0^2}\right)^\ga 
\end{equation}
where the exponents $\om$ and $\ga$ are related through the
characteristic function $\chi$
\begin{equation}\label{chi}
1\;=\;\frac{\asb}{\om}\;\chi(\ga) \,,
\;\;\;\;\;\;\;\;\;
\chi(\ga)= 2 \psi(1)-\psi(\ga)-\psi(1-\ga)
\,,
\end{equation}
with the QCD coupling $\as$ taken as a fixed parameter.\footnote{
The renormalisation group dependence of $\as$ on a scale is an
effect which goes beyond the leading order contribution.\protect\cite{NLO}}

For a general initial condition the asymptotic behaviour of
$\cF(x,k)$ at small $x$ is determined by
the leading singularity of $\ga\left({\asb\over\om}\right)$ in the $\om$-plane,
which is located at $\ga_c=\ga\big({\asb\over\om_c}\big)=\half$, giving
$\om_c=\asb\chi(\half)=4\asb\ln2$.

The analytic treatment of the CCFM equation is more complicated than
that of the BFKL equation because the gluon density contains one
extra parameter, $p$.
By analogy, we take the eigensolutions of \re{A1} in the form 
\begin{equation}
\label{ccfm-eigen}
x\cA(x,k,p) = 
x^{-\om} \frac{1}{k^2}\left(\frac{k^2}{k_0^2}\right)^{\tga} G \left( {p\over k} \right)
\,,
\end{equation}
where $\tga$ and $\om$ are related through
the unknown CCFM characteristic function $\tchi$
\begin{equation}\label{A5}
1=\frac{\asb}{\om}\tchi(\tga,\as)\,,
\end{equation}
and the function $G \left( {p\over k} \right)$
takes into account angular ordering, 
parameterising the unknown dependence on $p$.

For $0<\tga<1$ fixed, one obtains a coupled pair of equations for 
$G$ and $\tchi$:
\begin{equation}\label{dG}
p\partial_p\;G \left( {p\over k} \right)=
\asb
\int_p\;\frac{d^2\bq} {\pi q^2}
\left(\frac{p}{q}\right)^{\asb\tchi}\!\!\De\left({p\over q},q,k\right)
G\left(\frac{q}{\bkq}\right)
\left(\frac{\bkq^2}{k^2}\right)^{\tga-1}
\, ,
\end{equation}
with the initial condition $G(\infty)=1$, and
\begin{equation}
\tchi=\int\frac{d^2\bq}{\pi q^2}
\left[
\left( {\bkq^2\over k^2} \right)^{\tga-1}
\;  G \left( {q\over\bkq}\right)
-\theta(k-q)\; G\left({q\over k}\right) \right]
\,.
\end{equation}

By putting $G=1$ in this last equation, 
one notes that $\tchi$ becomes just the BFKL characteristic function \re{chi}. 
Since $1-G \left( {p\over k} \right)$ is formally of order $\as$, 
this demonstrates that angular ordering has a next-to-leading effect 
on structure function evolution, and one can also shows that the 
first corrections to the small-$x$ anomalous dimension are of the form
$\asb^3/\om^2$.

Though a number of asymptotic properties of the function
$G \left( {p\over k} \right)$ have been
determined,\cite{bmss} it has not so far been possible to obtain its
full analytic form. 

\section{Numerical results}

In this section we summarise the main results of the numerical analysis
carried out \cite{bmss} both for BFKL and CCFM equations in order to
gain further insight into the (subleading) effects of angular ordering
on the structure function.

\begin{figure}
\begin{center}
\epsfig{file=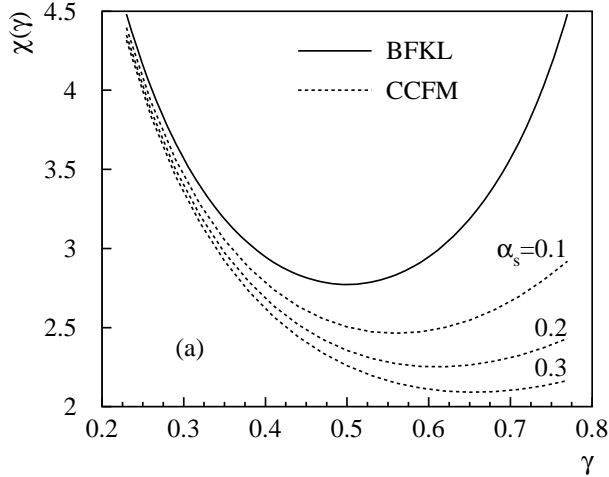,width=0.7\textwidth}
\end{center}
\caption[]{The characteristic functions with and without angular
 ordering; $\tchi(\ga,\as)$ and $\chi(\ga)$ are plotted as functions
 of $\ga$.} 
\label{fig:chi}
\end{figure}

Fig.~\ref{fig:chi} shows the results for $\tchi$ compared to the BFKL
characteristic functions as a function of $\tga$ for various $\as$.  
The difference $\de \chi=\chi-\tchi$ is positive, increases with
$\tga$, and increases with $\as$.  
Moreover we find $\de \chi \sim \tga$ for $\tga\to0$ ($\asb$
small and fixed) and $\de \chi \sim \asb$ for $\asb\to0$ ($\tga$
small and fixed). 
This implies that the next-to-leading correction to the
gluon anomalous dimension coming from angular ordering is of order
${\as^3\over\om^2}$.

\begin{figure}
\begin{center}
\epsfig{file=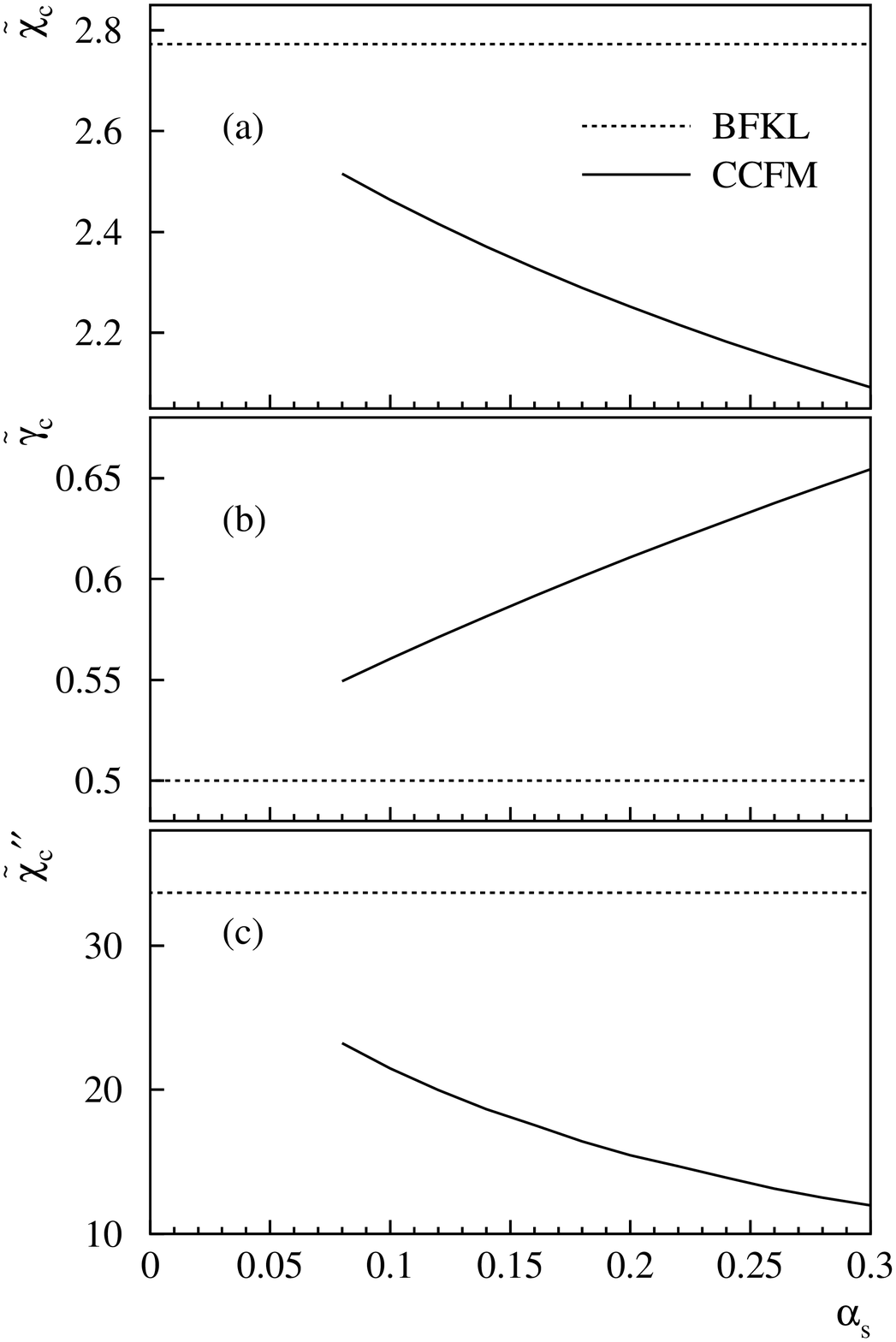,width=0.7\textwidth}\\
\end{center}
\caption[]{(a) The value of the minimum of the characteristic function,
 $\tchi_c$, as a function of $\as$. (b) The position of the minimum of
 the characteristic function, $\tga_c$, as a function of $\as$.
 (c) The second derivative of the characteristic function,
 ${\tchi_c}''$, at its minimum, as a function of $\as$.}
\label{fig:chiprop}
\end{figure}

With respect to the BFKL case, 
the position of the minimum of the characteristic function $\tchi$ 
gets shifted to the right, the value of the minimum is lowered
and --- in contrast to the BFKL case --- 
there is no longer even a divergence at $\ga=1$.
This behaviour of $\tchi$ reduces the exponent $\om_c$ of the small-$x$ growth
of the structure function, in accordance with the fact
that angular ordering reduces the phase space for evolution, 

In Fig.~\ref{fig:chiprop}a and \ref{fig:chiprop}b we plot as a 
function of $\as$ the values $\tchi_c$ and $\tga_c$
with $\tchi_c$ the minimum of $\tchi$ and $\tga_c$ its position.  
As expected the differences compared to the BFKL values $\chi_c=4\ln 2$
and $\ga_c=\half$ are of order $\asb$.  

Fig.~\ref{fig:chiprop}c shows the second derivative, ${\tchi_c}''$, of the
characteristic function at its minimum; this quantity is important
phenomenologically because the diffusion in $\ln k$ is inversely
proportional \footnote{This is strictly true only for 
the solution in the saddle-point approximation; 
nevertheless this quantity remains a good indicator,
due to the mild asymptotic behaviour of the $G$ function.}
to $\sqrt{{\tchi_c}''}$.
From this result, one can
therefore conclude that the inclusion of angular ordering
significantly reduces the diffusion compared to the BFKL case.

The loss of symmetry under $\ga\to1-\ga$ relates to the loss of
symmetry between small and large scales:
while in BFKL regions of small and large momenta are equally important,
in the CCFM case angular ordering favours instead the region of
larger $k$. 
However, at each intermediate branching, 
the region of vanishing momentum is still reachable for $x\to0$,  
so that the evolution still contains non-perturbative components.

\section{Final state distributions}

The inclusion of angular ordering is expected
to have relevant effects when
simple exclusive quantities, associated with 
one-gluon inclusive distributions, are considered.
Indeed, the cancellation between real emissions and virtual
corrections
--- which in the angular ordering equation for the inclusive structure
function reconstruct at leading level the BFKL solution --- 
is no longer guaranteed for the modified kernel which enter the
evolution equations for associated distribution. 

Although the analysis of this subject is far from being completed,
preliminary calculations confirm that both the shapes and the
normalisations of final state quantities are sensitive to
the phase spaces reduction associated with angular ordering. 

\begin{figure}
\begin{center}
\epsfig{file=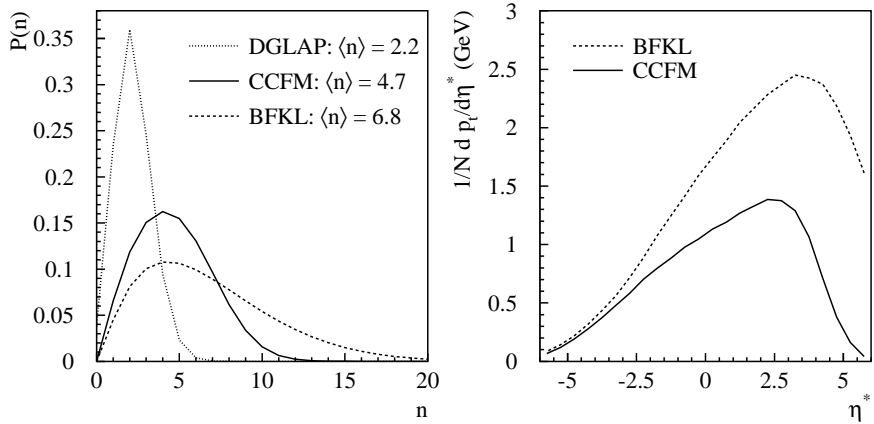,width=\textwidth}
\end{center}
\caption[]{(a) Distribution of number of emission with $q>q_0=1$GeV, for
 DGLAP, CCFM and BFKL evolution to $x = 5.10^{-5}$, $k=5\;$GeV, $\as=0.2$.
 (b) Transverse momentum flow in the hadronic centre of mass frame 
  as a function of the rapidity $\eta^*$
  for evolution to $x=2\cdot 10^{-4}$, $k=3\;\hbox{GeV}$, $\as=0.2$
  (the proton direction is to the left).}
\label{fig:figass}
\end{figure}

Fig.~\ref{fig:figass}a shows the distribution of the number 
of initial state gluons emitted.
As expected from the different behaviour in the collinear region, 
BFKL branching has more emissions and a broader tail
with respect to the CCFM case.

Fig.~\ref{fig:figass}b shows the $p_t$-distribution in rapidity.
As expected, angular ordering suppress the radiation in the central 
and high rapidity regions. 

\section*{Acknowledgements}

This research was carried out in collaboration with G.~Bottazzi,
G.~Marchesini and G.P.~Salam and supported in part by the Italian MURST.

\end{document}